# Does Flexoelectricity Drive Triboelectricity?

C. A. Mizzi[*], A. Y. W. Lin[*] and L. D. Marks[1]
Department of Materials Science and Engineering
Northwestern University, Evanston, IL 60208

[*]These authors contributed equally to this work

**Corresponding Author**

[1]To whom correspondence should be addressed. Email: L-marks@northwestern.edu.

**Abstract**

The triboelectric effect, charge transfer during sliding, is well established but the thermodynamic driver is not well understood. We hypothesize here that flexoelectric potential differences induced by inhomogeneous strains at nanoscale asperities drive tribocharge separation. Modelling single asperity elastic contacts suggests that nanoscale flexoelectric potential differences of ±1-10 V or larger arise during indentation and pull-off. This hypothesis agrees with several experimental observations, including bipolar charging during stick-slip, inhomogeneous tribocharge patterns, charging between similar materials, and surface charge density measurements.

The triboelectric effect, the transfer of charge associated with rubbing or contacting two materials, has been known for at least twenty-five centuries [1,2]. The consequences of this transfer are known to be beneficial and detrimental; for instance, tribocharging is widely exploited in technologies such as laser printers but can also cause electrostatic discharges that lead to fires. It is accepted that it involves the transfer of charged species, either electrons [3-5], ions [6,7], or charged molecular fragments [8], between two materials. The nature and identification of these charged species has been the focus of considerable research [2,9], but an important unresolved issue is the thermodynamic driver for charge transfer; the process of separating and transferring charge must reduce the free energy of the system. What is the charge transfer driver? In some cases specific drivers are well understood. For instance, when two metals with different work functions are brought into contact charge transfer will occur until the chemical potential of the electrons (Fermi level) is the same everywhere. Triboelectric charge transfer in insulators is less understood; proposed models include local heating [10] and trapped charge tunneling [11-13] but these models do not explicitly address the significant mechanical deformations associated with bringing two materials into contact and rubbing them together. Furthermore there is currently little ab-initio or direct numerical connection between experimental measurements and proposed drivers.

Since the pioneering work of Bowden and Tabor [14] it has been known that friction and wear at the nanoscale is associated with adhesion between, as well as the elastic and plastic deformation of, a statistical population of asperities. It is also well established that elastic deformation is thermodynamically linked to polarization: the linear coupling between strain and polarization is the piezoelectric effect and the linear coupling between strain gradient and polarization is the flexoelectric effect [15-17]. While piezoelectric contributions only occur for materials without an inversion center, flexoelectric contributions occur in all insulators and can be large at the nanoscale due to the intrinsic size scaling of strain gradients [17-19]. Quite a few papers have analyzed the implications of these coupling terms in phenomena including nanoindentation [20,21], fracture [22], and tunneling [23]. There also exists literature where the consequences of charging on friction have been studied [24-26], and frictional properties have been related to redistributions of interfacial charge density via first principles calculations [27]. However, triboelectricity, flexoelectricity, and friction during sliding are typically considered as three independent phenomena.

Are they really uncoupled phenomena? In this paper we hypothesize that the electric fields induced by inhomogeneous deformations at asperities via the flexoelectric effect lead to significant surface potentials differences, which can act as the driver for triboelectric charge separation and transfer. The flexoelectric effect may therefore be a very significant, and perhaps even the dominant, thermodynamic driver underlying triboelectric phenomena in many cases. To investigate this hypothesis in detail we analyze, within the conventional Hertzian [28] and Johnson-Kendall-Roberts (JKR) [29] contact models, the typical surface potential differences around an asperity in contact with a surface during indentation and pull-off. We find that surface potential differences in the range of ±1-10 V or more can be readily induced for typical polymers and ceramics at the nanoscale, and that the intrinsic asymmetry of the inhomogeneous strains during indentation and pull-off changes the sign of the surface potential difference. We argue that our model is consistent with a range of experimental observations, in particular bipolar tribocurrents associated with stick-slip [30], the scaling of tribocurrent with indentation force [31], the phenomenon of tribocharging of similar materials [32-35], and the inhomogeneous charging of insulators [36,37]. Taking the analysis a step further, our model suggests a suitable upper bound for the triboelectric surface charge density is the flexoelectric polarization that is found to be in semi-quantitative agreement with published experimental data without the need to invoke any empirical parameters. Given the recent ab-initio developments of flexoelectric theory [38-41], we argue that flexoelectricity can provide an ab-initio understanding of many triboelectric phenomena.

Nanoscale asperity contact consists of two main phenomena, indentation and pull-off, which are illustrated in Fig. 1. To investigate the electric fields arising from the strain gradients associated with these two processes, we combine the constitutive flexoelectric equations with the classic Hertzian and JKR models, for simplicity considering only vertical relative displacements; see later for some comments about shear. As discussed further in the Supplemental Material [57], the normal component of the electric field induced by a flexoelectric coupling in an isotropic non-piezoelectric half plane oriented normal to $\hat{z}$ is given by:

$$E_z = -f \frac{\partial \epsilon}{\partial z}\bigg|_{eff} = -f\big(3\epsilon_{zzz} + \epsilon_{zxx} + \epsilon_{xzx} + \epsilon_{xxz} + \epsilon_{zyy} + \epsilon_{yzy} + \epsilon_{yyz}\big) \qquad (1)$$

where $E_z$ is the electric field linearly induced by $\frac{\partial \epsilon}{\partial z}\big|_{eff}$ the effective strain gradient. The proportionality constant $f$ is the flexocoupling voltage (i.e., the flexoelectric coefficient divided by the dielectric constant) and the effective strain gradient is the sum of the symmetry-allowed strain gradient components (where $\epsilon_{jkl} = \frac{\partial \epsilon_{jk}}{\partial x_l}$).

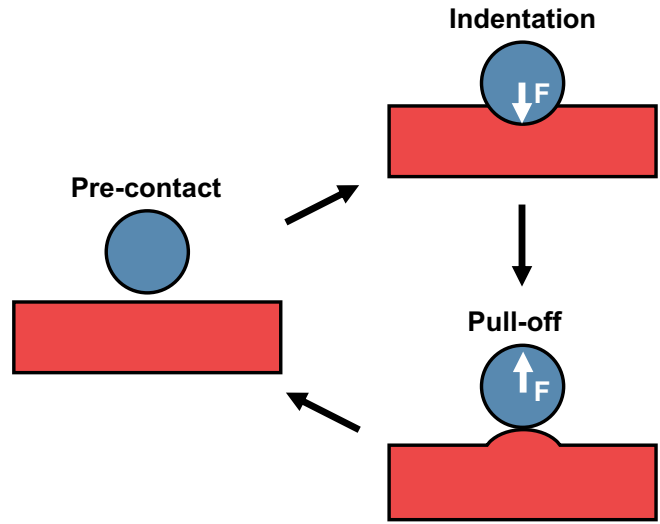

Fig. 1. Schematic of asperity contact between a rigid sphere (blue) and an elastic body (red). During indentation and pull-off the elastic body will deform, developing a net strain gradient opposite to the direction of the applied force (F).

First, we will analyze the indentation case. Because of the axial symmetry of Hertzian indentation, only five strain gradient components in Equation (1) are symmetrically inequivalent. Expressions for these components are derived from classic Hertzian stresses (see Supplemental Material [57]) and depicted in Fig. 2(a)-(e) as contour plots. From these plots it is evident that the strain gradient components have complex spatial distributions, the details of which depend on the materials properties of the deformed body (Young's modulus, Poisson's ratio) as well as external parameters (applied force, indenter size). Further insight can be gained by calculating the average effective strain gradient within the indentation volume, which is taken to be the cube of the deformation radius. The average effective strain gradient is negative and scales inversely with indenter size, independent of the materials properties of the deformed body and the applied force. The former is intuitive since a material deformed by an indenter should develop a curvature opposite to the direction of the applied force, and the latter is a consequence of averaging (Supplemental Material [57]). As shown in Fig. 2(f), the average effective strain gradient associated with Hertzian indentation is on the order of $-10^8$ m$^{-1}$ in all materials at the nanoscale. Such large strain gradients immediately suggest the importance of flexoelectric couplings [17,18].

For pull-off we use JKR theory, which incorporates adhesion effects between a spherical indenter and an elastic half-space into the Hertz contact model. The tensile force required to separate the indenter from the surface, also known as the pull-off force, can be written as

$$F_{adh} = -\frac{3}{2}\pi\Delta\gamma R \tag{2}$$

where $\Delta\gamma$ is the adhesive energy per unit area and $R$ is the radius of the spherical indenter. Replacing the applied force in the Hertzian indentation strain gradient expressions with this force yields pull-off strain gradients immediately before contact is broken. This analysis for the pull-off case yields strain gradient distributions qualitatively similar to those shown in Fig. 2, except with opposite signs because the force is applied in the opposite direction. Importantly, as in the indentation case, the average effective strain gradient within the pull-off volume scales inversely with indenter size, is independent of the materials properties of the deformed body, and is on the order of $10^8$ m$^{-1}$ in all materials at the nanoscale.

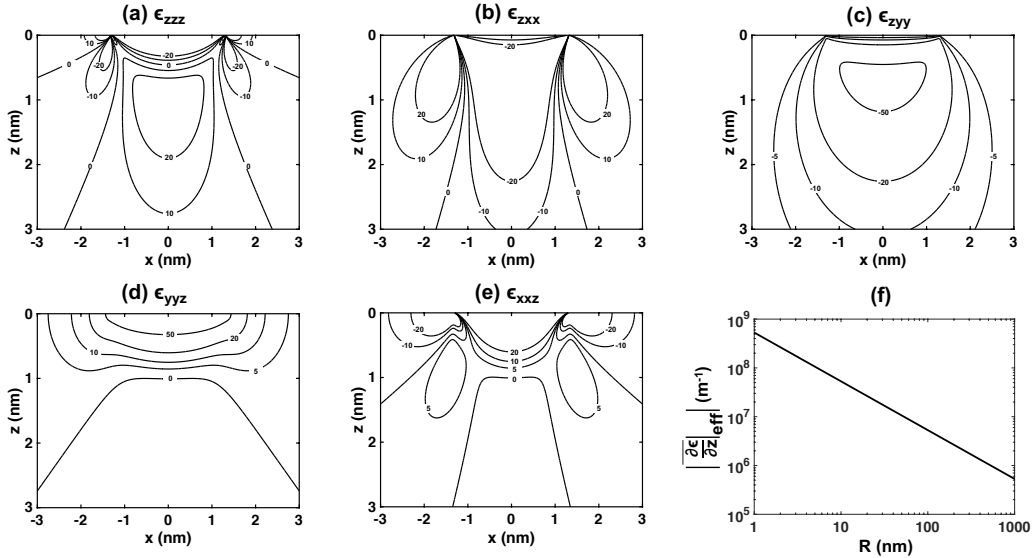

Fig. 2. (a) – (e) Symmetrically inequivalent strain gradients arising from Hertzian indentation of an elastic half-space that can flexoelectrically couple to the normal component of the electric field.

Lines indicate constant strain gradient contours in units of $10^6$ m$^{-1}$, $z$ is the direction normal to the surface with positive values going into the bulk, $x$ is an in-plane direction, and the origin is the central point of contact. Data corresponds to 1 nN of force (a conservatively small number) applied to an elastic half-space with a Young's modulus of 3 GPa and a Poisson's ratio of 0.3 (typical polymer) by a 10 nm rigid indenter. (f) The magnitude of the average effective strain gradient ($\left|\overline{\frac{\partial \epsilon}{\partial z}}\right|_{eff}$) as a function of indenter radius ($R$). The average effective strain gradient corresponds to a sum of the strain gradient components shown in (a) – (e) averaged over the indentation/pull-off volumes.

We now turn to the flexoelectric response to these deformations. Obtaining analytical expressions for the normal component of the electric field in the deformed body induced by indentation and pull-off involves substituting the strain gradient components shown in Fig. 2 into Equation (1). This electric field component is shown in Fig. 3 for the indentation case with a positive flexocoupling voltage. The pull-off case is similar, but the signs of the electric fields are reversed. Because the electric field induced by the flexoelectric effect is the effective strain gradient scaled by the flexocoupling voltage, its magnitude is linearly proportional to the flexocoupling voltage and inversely proportional to the indenter size. The average electric field in the indentation/pull-off volume is on the order of $10^8$-$10^9$ V/m for all materials at the nanoscale assuming a conservative flexocoupling voltage of 1 V [16,17,42]; some specific flexocoupling voltages are given in Supplemental Tables S1 and S2 [57].

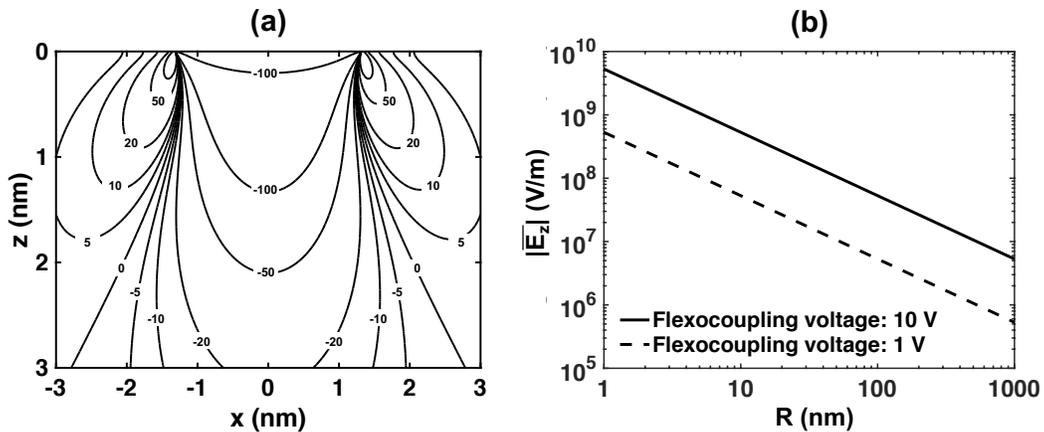

Fig. 3. (a) Normal component of the electric field induced by Hertzian indentation via a flexoelectric coupling. Lines indicate constant electric field contours in units of MV/m, $z$ is the direction normal to the surface with positive values going into the bulk, $x$ is an in-plane direction, and the origin is the central point of contact. Data corresponds to 1 nN of force applied to an elastic half-space with a Young's modulus of 3 GPa and a Poisson's ratio of 0.3 (typical polymer) by a 10 nm indenter. A flexocoupling voltage of 1 V is assumed. (b) Magnitude of the average electric field ($|\overline{E_z}|$) in the indentation/pull-off volumes as a function of indenter radius ($R$) assuming a flexocoupling voltage of 1 V (dashed) and 10 V (solid).

The electric fields induced by the flexoelectric effect in the bulk of the deformed body will generate a potential on its surface. Figure 4 depicts the surface potential difference calculated from the normal component of the electric field (Supplemental Material [57]) along the deformed surface of a typical polymer with a flexocoupling voltage of 10 V [16,17,42]; the available measured flexocoupling voltages for polymers indicates that this may be a significant underestimate, see Supplemental Table S2 [57]. The pull-off surface potential difference tends to be larger in magnitude and spatial extent than the indentation surface potential difference. In both cases the magnitude of the maximum surface potential difference is sensitive to the materials properties of the deformed body (Young's modulus, Poisson's ratio, adhesion energy, flexocoupling voltage) and external parameters (applied force, indenter size). Specifically, the surface potential differences for indentation and pull-off scale as

$$V_{indentation,min} \propto -f \left(\frac{F}{R^2 Y}\right)^{1/3} \tag{3}$$

$$V_{pull-off,max} \propto f \left(\frac{\Delta \gamma}{R\, Y}\right)^{1/3} \tag{4}$$

where $V_{indentation,min}$ is the minimum surface potential difference for indentation, $V_{pull-off,max}$ is the maximum surface potential difference for pull-off, $f$ is the flexocoupling voltage, $F$ is the applied force, $R$ is the indenter radius, $Y$ is the Young's modulus, and $\Delta \gamma$ is the energy of adhesion.

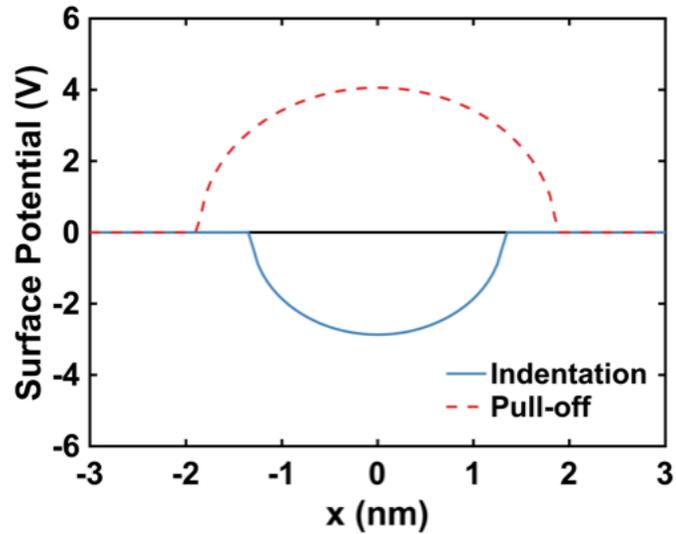

Fig. 4. Electric potential difference along the surface of the deformed body for indentation (solid) and pull-off (dashed). $x$ is an in-plane direction and the origin is the central point of contact. Data corresponds to 1 nN of force applied to an elastic half-space with a Young's modulus of 3 GPa, a Poisson's ratio of 0.3, adhesion energy of 0.06 N/m (typical polymer), and flexocoupling voltage of 10 V by a 10 nm indenter.

The above analysis indicates that large strain gradients arising from deformations by nanoscale asperities yield surface potential differences via a flexoelectric coupling in the ±1-10 V range, as a conservative estimate. The magnitude of this surface potential difference is sufficient to drive charge transfer, suggesting that flexoelectric couplings during indentation and pull-off can be responsible for triboelectric charging. Furthermore, this model implies that the direction of charge transfer is controlled by a combination of the direction of the applied force and local topography (i.e. is the asperity indenting or pulling-off), as well as the sign of the flexocoupling voltage.

These features are consistent with and can explain a significant number of previous triboelectric observations without introducing any adjustable parameters. First, it has been observed that tribocurrents exhibit bipolar characteristics associated with stick-slip [30]. This bipolar nature is consistent with the change in the sign of the surface potential difference for indentation and pull-off predicted by our model. We note that these experiments had some shear component which is not exactly the same as our analysis and complicates the problem due to the breakdown of circular symmetry. While this will yield a more complex strain gradient distribution than our simplified model, the total potential difference will be the sum of normal and shear contributions which does not change our general conclusions. Second, the tribocurrent has been shown to scale with the indentation force to the power of $\frac{1}{3}$ [31], which matches the scaling of the indentation surface potential difference with force. Thirdly, charging between similar materials [32-35] and the formation of non-uniform tribocharge patterns [36,37,43,44] can be explained by considering the effect of local surface topography and crystallography on the direction of charge transfer: local variation in surface topography dictates which material locally acts as the asperity, and consequently the direction in which charge transfers. In addition, it is established for crystalline materials that both the magnitude and sign of the flexocoupling voltage can change with crystallographic orientation (Supplemental Table S1 [57]). Finally, recent work has demonstrated that macroscopic curvature biases tribocharging so that convex samples tend to charge negative and concave samples tend to charge positive; this coupling between curvature and charge transfer direction is a natural consequence of our flexoelectric model [45].

Going beyond these qualitative conclusions, it is relevant to explore whether flexoelectricity can quantitatively explain experimental triboelectric charge transfer measurements. An important quantitative parameter in the triboelectric literature is the magnitude of triboelectric surface charge density which has been measured in a number of systems including spherical particles [35,46] and patterned triboelectric devices [47,48], and normally enters models as an empirical parameter [49,50]. We hypothesize that the upper bound for the triboelectric surface charge density is set by the flexoelectric polarization, i.e. charge will transfer until the flexoelectric polarization is screened (Supplemental Material [57]). As shown in Table 1, this hypothesis agrees with existing tribocharge measurements on a range of length scales to within an order of magnitude without invoking anomalous flexoelectric coefficients.

| Reference | Feature Size | $\sigma_{tribo}$ ($\mu C/m^2$) | $P_{FxE}$ ($\mu C/m^2$) |
|---|---|---|---|
| [46] | 2.8 mm | 0.5 | 0.4 |
| [35] | 326 μm | 0.2 | 1.6 |
| [35] | 251 μm | 0.5 | 2.1 |
| [47,48] | 10 μm | 97.4 | 106.1 |

Table 1. Comparison between measured triboelectric surface charge ($\sigma_{tribo}$) and calculated flexoelectric polarization ($P_{FxE}$) for feature sizes in the mm to μm range assuming a flexoelectric coefficient of 1 nC/m.

These results make a strong case that the flexoelectric effect drives triboelectric charge separation and transfer, and that nanoscale friction, flexoelectricity, and triboelectricity occur simultaneously and are intimately linked: macroscopic forces during sliding on insulators cause local inhomogeneous strains at contacting asperities which induce significant local electric fields which in turn drive charge separation. This analysis does not depend upon the details of the charge species, they may be electrons, polymeric ions, charged point defects in oxides, or some combination. Hence our model does not contradict any of the existing literature on the nature of the charge species, instead it provides a thermodynamic rationale for the charge separation to occur. We have deliberately used very conservative numbers for the flexocoupling voltage, and many materials are known to have significantly larger values – see Supplemental Tables S1 and S2 [57]. It is therefore very plausible that much larger potential differences can be generated. Our analysis also suggests ways to optimize charge separation (e.g. assuming pull-off dominates, based upon Equation (4) one wants a relatively soft material with a high flexocoupling voltage, large adhesion, and many small asperities). Some additional experimental and theoretical ways to assess this model are discussed briefly in the Supplemental Material [57].

In addition, the formalism we have used is not limited to inorganic materials, but is quite general. As one extension it is known that semi-crystalline layers are formed at the confined spaces during sliding in a lubricant [51], so it is not unreasonable that flexoelectric effects can drive charge separation in lubricants. Another extension is biological materials, as flexoelectric effects in biological membranes are well-established [52]. We also note the magnitude of the flexoelectricity-induced electric fields and surface potential differences at asperities (and crack tips [22]) suggest flexoelectricity can play a role in triboluminescence [53-55], triboplasma generation [56] or tribochemical reactions. Such hypotheses merit further work.

In summary, using the Hertz and JKR models for indentation and pull-off, we show that deformations by nanoscale asperities yield surface potential differences via a flexoelectric coupling in the $\pm 1$-10 V range or more, large enough to drive charge separation and transfer. The direction and magnitude of the surface potential differences depend on the applied force, asperity size, local topography, and material properties. These findings explain some previous tribocharging observations and we argue are the first steps towards an ab-initio understanding of triboelectric phenomena.

**Acknowledgements:** This work was supported by the National Science Foundation Grant No. CMMI-1400618 (AYWL) and the U.S. Department of Energy, Office of Science, Basic Energy Sciences, under Award No. DE-FG02-01ER45945 (CAM).

**Author Contributions:** CAM and AYWL performed the analysis supervised by LDM. All authors contributed to the writing of the Letter.

# Supplemental Material: Does Flexoelectricity Drive Triboelectricity?

C. A. Mizzi[*], A. Y. W. Lin[*] and L. D. Marks[1]

Department of Materials Science and Engineering

Northwestern University, Evanston, IL 60208

[*]These authors contributed equally to this work

**Corresponding Author**

[1]To whom correspondence should be addressed. Email: L-marks@northwestern.edu.

## S1. General details

Flexoelectricity, tribology, and triboelectricity are separate subdisciplines, each with their own terminology and literature. Since this paper discusses the cross-connections between the three some general discussion of flexoelectricity and tribology is provided here for readers who are less familiar with them.

The flexoelectric effect describes the linear coupling between polarization and strain gradient. Because flexoelectricity relates strain gradient (a third-rank tensor) and polarization (a first-rank tensor), the flexoelectric effect is a fourth-rank tensor property described by a coupling coefficient known as the flexoelectric coefficient. Flexoelectricity was first observed in solids by Bursian and Zaikovskii [1], but the term flexoelectricity was coined in the field of liquid crystals [2]. Flexoelectric characterization of relaxor ferroelectrics by Cross and Ma in the early 2000s [3] rekindled interest in this subject in the context of oxides, and since then there has been significant experimental [4-7] and theoretical [8-14] progress in understanding flexoelectricity.

While piezoelectric effects are well established, they only occur for crystallographies which do not have an inversion center. In contrast, flexoelectric effects occur in all insulators independent of the crystallography. The coupling is a fourth-order tensor which can conveniently be scaled by the dielectric coefficient; this is known as the flexocoupling voltage. For many materials flexocoupling voltages are in the range of 1-10 V, although there are cases where it is significantly larger, for instance ~40 V for $DyScO_3$ [15]. Both the sign and the magnitude of the flexocoupling voltage in a given material can depend strongly upon the crystallographic orientation [16], and a few experimental values are given in Table S1 for single crystal oxides and

Table S2 for polymers. It is known that in some cases the effective flexocoupling voltage can be anomalously large if there is coupling to charged defects [4] in the material or other effects [17]. This is an area of continuing research. Some recent reviews can be found in references [3,16,18].

| Reference | Material | Orientation | µ (nC/m) | f (V) |
|---|---|---|---|---|
| [5] | $SrTiO_3$ | [001] | 6.1 | 2.3 |
| [5] | $SrTiO_3$ | [101] | -5.1 | -1.9 |
| [5] | $SrTiO_3$ | [111] | -2.4 | -0.9 |
| [19] | $BaTiO_3$ | [001] | 200 | 22 |
| [19] | $BaTiO_3$ | [110] | -50 | -6 |
| [19] | $BaTiO_3$ | [111] | -10 | -2 |
| [15] | $DyScO_3$ | [110] | -8.4 | -42 |
| [4] | $TiO_2$ | [001] | 2 | 1.3 |
| [20] | PIN-PMN-PT | [001] | ~ 4e4 | ~ 1400 |

Table S1. A few examples illustrating how the magnitude and sign of the flexoelectric coefficient change with crystallographic orientation in oxide single crystals. All measurements were made at room temperature in a three-point bending geometry with orientation corresponding to the bending direction. Note, the flexoelectric coefficient reported here is a linear combination of tensor coefficients. For more details, see [4, 5, 15, 16, 19, 20].

| Reference | Material | \|μ\| (nC/m) | \|f\| (V) |
|---|---|---|---|
| [21] | PVDF | 13 | 160 |
| [21] | Oriented PET | 9.9 | 289 |
| [21] | Polyethylene | 5.8 | 273 |
| [21] | Epoxy | 2.9 | 84 |
| [22] | P(VDF-TrFE) (70/30) | 3.0 | 31 |
| [22] | P(VDF-TrFE) (55/45) | 4.2 | 43 |
| [22] | P(VDF-TrFE-CTFE) | 3.5 | 10 |
| [23] | Jeffamine | ~ 0.1 | ~ 4.5 |
| [23] | Keltan | ~ 0.1 | ~ 4.5 |
| [23] | Natural Rubber | ~ 0.5 | ~ 20 |

Table S2. Some examples illustrating the magnitude of the flexoelectric coefficient and flexocoupling voltage in polymers. Measurements from Ref. [21] were made in a cantilever geometry, measurements from Ref. [22] were made in a three-point bending geometry, and measurements from Ref. [23] were made in a truncated pyramid geometry. The signs of the coefficients were not specified. For more details, see [16, 21, 22, 23].

While the classic continuum laws of friction have been known for centuries, since the pioneering work of Bowden and Tabor it has been established that they are statistical averages over many asperity contacts. In many cases, single asperities do not follow the statistical macroscopic laws of friction, as reviewed in [24]. While there remains some debate about the exact mechanisms of energy dissipation in sliding contact, for instance the importance of electron coupling [25] versus movement of misfit dislocations [26,27], collective motion of dislocations [28,29] or local chemistry [30,31], the general nature is well understood. The consequences of tribocharging on increasing friction has also been explored in the literature [32-34].

## S2. Flexoelectricity in an isotropic non-piezoelectric material

The constitutive equation for flexoelectricity in a non-piezoelectric dielectric material is

$$D_i = \mu_{ijkl} \frac{\partial \epsilon_{jk}}{\partial x_l} + K_{ij} E_j \quad (S1)$$

where $D_i$ is the dielectric displacement, $\mu_{ijkl}$ is the flexoelectric coefficient, $\frac{\partial \epsilon_{jk}}{\partial x_l}$ is the (symmetrized) strain gradient, $K_{ij}$ is the dielectric constant, $E_j$ is the electric field, and subscripts are Cartesian directions using the Einstein convention. In this work, we will assume the material is isotropic which greatly reduces the number of non-trivial components of $\mu_{ijkl}$ and $K_{ij}$.

$$\mu_{ijkl} = \alpha \delta_{ij}\delta_{kl} + \beta \delta_{ik}\delta_{jl} + \gamma \delta_{il}\delta_{jk} \quad (S2)$$

$$K_{ij} = K\delta_{ij} \quad (S3)$$

Additionally, we assume the non-trivial components of the isotropic flexoelectric coefficient tensor are approximately the same so

$$\mu_{ijkl} = \mu \left( \delta_{ij}\delta_{kl} + \delta_{ik}\delta_{jl} + \delta_{il}\delta_{jk} \right) \quad (S4)$$

In the absence of surface charge the normal component of the dielectric displacement vanishes.

$$\sigma = \hat{\boldsymbol{n}} \cdot \vec{\boldsymbol{D}} = 0 \quad (S5)$$

Taking $\hat{\boldsymbol{n}} = \hat{\boldsymbol{z}}$ and combining the surface charge condition with the constitutive equation for a non-piezoelectric, isotropic dielectric material yields an expression for the normal component of the electric field induced by a flexoelectric coupling.

$$E_z = -f \left.\frac{\partial \epsilon}{\partial z}\right|_{eff} = -f\left(3\epsilon_{zzz} + \epsilon_{zxx} + \epsilon_{xzx} + \epsilon_{xxz} + \epsilon_{zyy} + \epsilon_{yzy} + \epsilon_{yyz}\right) \quad (S6)$$

In this expression, the symmetry-allowed strain gradients are denoted explicitly ($\frac{\partial \epsilon_{jk}}{\partial x_l} = \epsilon_{jkl}$) and the flexoelectric coefficient normalized by the dielectric constant has been replaced by the flexocoupling voltage $f$.

### S3. Hertzian strain gradients from Hertzian stresses

For a non-adhesive single asperity contact, the Hertzian contact model [35] is used to describe the deformation mechanics of a rigid sphere on an elastic flat half-space. The Hertzian model assumes the materials in contact to be homogeneous and isotropic, and deformations are perfectly elastic and governed by classical continuum mechanics (Hooke's law). Furthermore, the Young's modulus $Y$ and Poisson's ratio $v$ are also assumed to be constant under load.

The stress fields given by the Hertzian contact model with a spherical indenter have been thoroughly analyzed elsewhere [36], so we will merely use the results of this model to derive analytical expressions for the Hertzian strain gradient fields. In cylindrical coordinates, stresses in the bulk of a Hertzian deformed elastic half-space are given by:

$$\frac{\sigma_{rr}}{p_m} = \frac{3}{2}\left\{\frac{1-2\nu}{3}\frac{a^2}{r^2}\left(1-\left(\frac{z}{u^{1/2}}\right)^3\right) + \left(\frac{z}{u^{1/2}}\right)^3 \frac{a^2 u}{u^2 + a^2 z^2}\right.$$
$$\left. + \frac{z}{u^{1/2}}\left(u\frac{1-\nu}{a^2+u} + (1+\nu)\frac{u^{1/2}}{a}\tan^{-1}\left(\frac{a}{u^{1/2}}\right) - 2\right)\right\} \quad (S7)$$

$$\frac{\sigma_{\theta\theta}}{p_m} = -\frac{3}{2}\left\{\frac{1-2\nu}{3}\frac{a^2}{r^2}\left(1-\left(\frac{z}{u^{\frac{1}{2}}}\right)^3\right) + \frac{z}{u^{\frac{1}{2}}}\left(2\nu + u\frac{1-\nu}{a^2+u} + (1+\nu)\frac{u^{\frac{1}{2}}}{a}\tan^{-1}\left(\frac{a}{u^{\frac{1}{2}}}\right)\right)\right\} \quad (S8)$$

$$\frac{\sigma_{zz}}{p_m} = -\frac{3}{2}\left\{\left(\frac{z}{u^{\frac{1}{2}}}\right)^3 \frac{a^2 u}{u^2 + a^2 z^2}\right\} \quad (S9)$$

$$\frac{\sigma_{rz}}{p_m} = -\frac{3}{2}\left(\frac{r z^2}{u^2 + a^2 z^2}\right)\left(\frac{a^2 u^{\frac{1}{2}}}{a^2 + u}\right) \quad (S10)$$

where

$$u = \frac{1}{2}\{(r^2 + z^2 - a^2) + ((r^2 + z^2 - a^2)^2 + 4a^2 z^2)^{1/2}\} \quad (S11)$$

$$p_m = \frac{F}{\pi a^2} \quad (S12)$$

$$a = \left(\frac{3}{4}\frac{F R}{Y}(1-\nu^2)\right)^{1/3} \quad (S13)$$

In these expressions, $F$ is the applied force, $a$ is the deformation radius, $Y$ is the Young's modulus, $\nu$ is the Poisson's ratio, and $r$ and $z$ are cylindrical coordinates. There is no $\theta$ dependence is these formulas because of radial symmetry.

These stresses were related to strain using the isotropic Hooke's law in cylindrical coordinates:

$$\epsilon_{rr} = \frac{1}{Y}(\sigma_{rr} - \nu(\sigma_{\theta\theta} + \sigma_{zz})) \quad (S14)$$

$$\epsilon_{\theta\theta} = \frac{1}{Y}(\sigma_{\theta\theta} - \nu(\sigma_{rr} + \sigma_{zz})) \quad (S15)$$

$$\epsilon_{zz} = \frac{1}{Y}\left(\sigma_{zz} - v(\sigma_{\theta\theta} + \sigma_{rr})\right) \quad (S16)$$

$$\epsilon_{rz} = \frac{2(1+v)}{Y}\sigma_{rz} \quad (S17)$$

Then, these cylindrical strains were transformed into Cartesian strains using the transformation matrix

$$T = \begin{pmatrix} \cos(\theta) & -\sin(\theta) & 0 \\ \sin(\theta) & \cos(\theta) & 0 \\ 0 & 0 & 1 \end{pmatrix} \quad (S18)$$

Lastly, these expressions were differentiated to determine expressions for relevant strain gradient components. The strain gradients that can couple to the normal component of the electric field as derived in S1 are $\epsilon_{zzz}$, $\epsilon_{zxx}$, $\epsilon_{xxz}$, $\epsilon_{xzx}$, $\epsilon_{zyy}$, $\epsilon_{yyz}$, and $\epsilon_{yzy}$. Each of these strain gradients are functions of $r, z, R, F, Y,$ and $v$. Additionally, because of the axial symmetry of this problem, $\epsilon_{yzy} = \epsilon_{zyy}$ and $\epsilon_{xzx} = \epsilon_{zxx}$.

## S4. Pull-off Model

The adhesion between contacts has been extensively studied by two groups: Johnson, Kendall, and Roberts (JKR) [37] and Derjaguin, Muller, and Toporov (DMT) [38]. The JKR theory, which incorporates adhesion effects due to the increased contact area caused by the elastic surface into the Hertzian contact model, is used to model pull-off. In the JKR theory, long-range adhesive interactions outside the contact area are neglected so it has been shown that this model is appropriate for systems with compliant materials that have high adhesion and large indenter radii [39]. Additionally, the JKR theory will also predict a non-zero contact area even in the absence of applied loads, resulting in a tensile load required to separate the two adhered surfaces. The pull-off force is the maximum magnitude of this tensile load.

In contrast, the DMT theory includes adhesion interactions outside the contact area by considering long-range adhesive interactions such as van der Waals forces. Consequently, the DMT theory is better for approximating systems with stiff materials that have weak, long-range adhesion and small indenter radii. In reality, most physical systems will fall between the JKR and DMT limits, which are described quantitatively by Tabor's parameter $\mu_T$ [39]:

$$\mu_T = \left(\frac{R\Delta\gamma(1-v^2)^2}{Y^2 z_o^3}\right)^{\frac{1}{3}} \quad (S19)$$

where $R$ is the radius of the spherical indenter, $\Delta\gamma$ is the adhesive energy per unit area, $v$ is the Poisson's ratio, $Y$ is the Young's modulus, and $z_o$ is the equilibrium separation distance. Using the materials properties for a typical polymer ($Y$ = 3 GPa, $v$ = 0.3, $\Delta\gamma$ = 0.06 N/m) and setting $z_o$ to the bond length of a C-C bond (1.54 Å), $\mu_T$ is calculated as 0.97, which is in between the JKR and DMT limits but closer to the JKR limit. For typical ceramics ($Y$ = 250 GPa, $v$ = 0.3, $\Delta\gamma$ = 2 N/m), $\mu_T$ is calculated to be 0.53, showing that the JKR theory may not yield the best approximation. However, since we are not using the full JKR theory to model deformations and their induced polarizations and only using the pull-off force expression, this analysis demonstrates that the adhesive tensile loads predicted by the JKR theory is sufficient for the purposes of this Letter.

## S5. Average strain gradient for indentation and pull-off

Recall the effective strain gradient, defined as

$$\left.\frac{\partial\epsilon}{\partial z}\right|_{eff} = 3\epsilon_{zzz} + \epsilon_{zxx} + \epsilon_{xzx} + \epsilon_{xxz} + \epsilon_{zyy} + \epsilon_{yzy} + \epsilon_{yyz} \quad (S20)$$

in the main text, linearly induces an electric field via the flexoelectric effect. It is comprised of a number of strain gradient components, each with complex spatial distributions. Therefore, to get a sense of the overall magnitude and impact of the effective strain gradient, it is convenient to average it. A natural choice of integration volume is the deformation volume defined as $a^3$, where $a$ is the deformation radius.

$$\overline{\left.\frac{\partial\epsilon}{\partial z}\right|_{eff}} = \frac{1}{a^3}\int \left.\frac{\partial\epsilon}{\partial z}\right|_{eff} dV \quad (S21)$$

This is particularly convenient because for indentation, $\left.\frac{\partial\epsilon}{\partial z}\right|_{eff}$ is a function of materials parameters and the applied force via $a$. Similarly, for pull-off $\left.\frac{\partial\epsilon}{\partial z}\right|_{eff}$ is a function of materials parameters via $a$. Therefore, averaging over the deformation volume effectively removes all dependences except for the indenter radius. This is confirmed numerically. Moreover, since

$$E_z = -f\left.\frac{\partial\epsilon}{\partial z}\right|_{eff} \quad (S22)$$

and $f$ is a constant, it follows that an average electric field can be defined as

$$\overline{E_z} = -f \left.\frac{\overline{\partial \epsilon}}{\partial z}\right|_{eff} \quad (S23)$$

which is also independent of materials properties and applied parameters except the indenter size.

## S6. Comparison between indentation and pull-off flexoelectric responses

To model the pull-off case, the Hertz expressions for the deformation radius and pressure are replaced with JKR expressions. Namely,

$$a_{Hertz} = \left(\frac{3}{4}\frac{F R}{Y}(1-v^2)\right)^{1/3} \quad (S24)$$

$$p_{Hertz} = \frac{F}{\pi a^2} \quad (S25)$$

$$a_{JKR} = \left(\frac{9\pi \Delta\gamma R^2}{8 Y}(1-v^2)\right)^{1/3} \quad (S26)$$

$$p_{JKR} = -\frac{3}{2}\frac{\Delta\gamma R}{a^2} \quad (S27)$$

The net effect on the induced electric field is demonstrated below in Fig. S1 for a typical polymer ($Y = 3$ GPa, $v = 0.3$, $\Delta\gamma = 0.06$ N/m, and $f = 10$ V) contacted by a rigid sphere with radius $R = 10$ nm and an indentation force $F = 1$ nN. This plot depicts the magnitude of the normal component of the electric field at the central point of contact ($x = 0$, $y = 0$) as a function of depth into the bulk of the deformed body ($z$). From this plot, it is apparent that besides the change in the sign, the main difference between the pull-off and indentation electric fields is their spatial distribution.

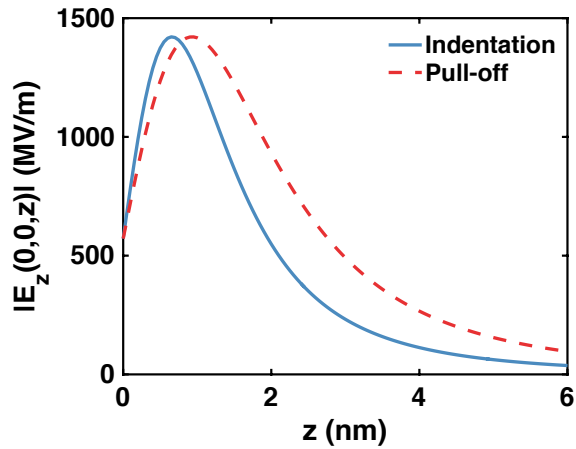

Fig. S1. Comparison between normal component of the electric field in the indentation and pull-off cases. Data corresponds to the normal component of the electric field at the central point of contact ($x = 0$, $y = 0$) as a function of depth into the bulk of the deformed body ($z$) for typical

polymer ($Y = 3$ GPa, $\nu = 0.3$, $\Delta\gamma = 0.06$ N/m, and $f = 10$ V) contacted by a rigid sphere with radius $R = 10$ nm and an indentation force $F = 1$ nN.

## S7. Surface potential difference: calculation and scaling relationships

The electric fields induced by the flexoelectric effect in the bulk of the deformed body will also generate a potential on its surface. This flexoelectric surface potential difference can be calculated from the normal component of the electric field via

$$V_{surface}(x, y) = -\int_{z=0}^{z=\infty} E_z(x, y, z) \, dz \quad (S28)$$

A convenient way to characterize the size of this surface potential difference is the magnitude of the minimum surface potential difference during indentation and maximum surface potential difference during pull-off. These values correspond to $V_{surface}(x = 0, y = 0)$.

The set of figures below demonstrate how the indentation and pull-off surface potential differences scale with materials properties and external parameters. They were obtained by calculating the indentation and pull-off surface potential differences while varying one property/parameter with all other terms held constant. Power-law fits used to determine the scaling behavior are shown in red in Fig. S2 and S3. The end results are summarized in the expressions

$$V_{indentation,min} \propto -f \left(\frac{F}{R^2 Y}\right)^{\frac{1}{3}} \quad (S29)$$

$$V_{pull-off,max} \propto f \left(\frac{\Delta\gamma}{R\,Y}\right)^{\frac{1}{3}} \quad (S30)$$

The surface potential differences above are also roughly linear with $(1 - \nu^2)$, but this proportionality is not exact.

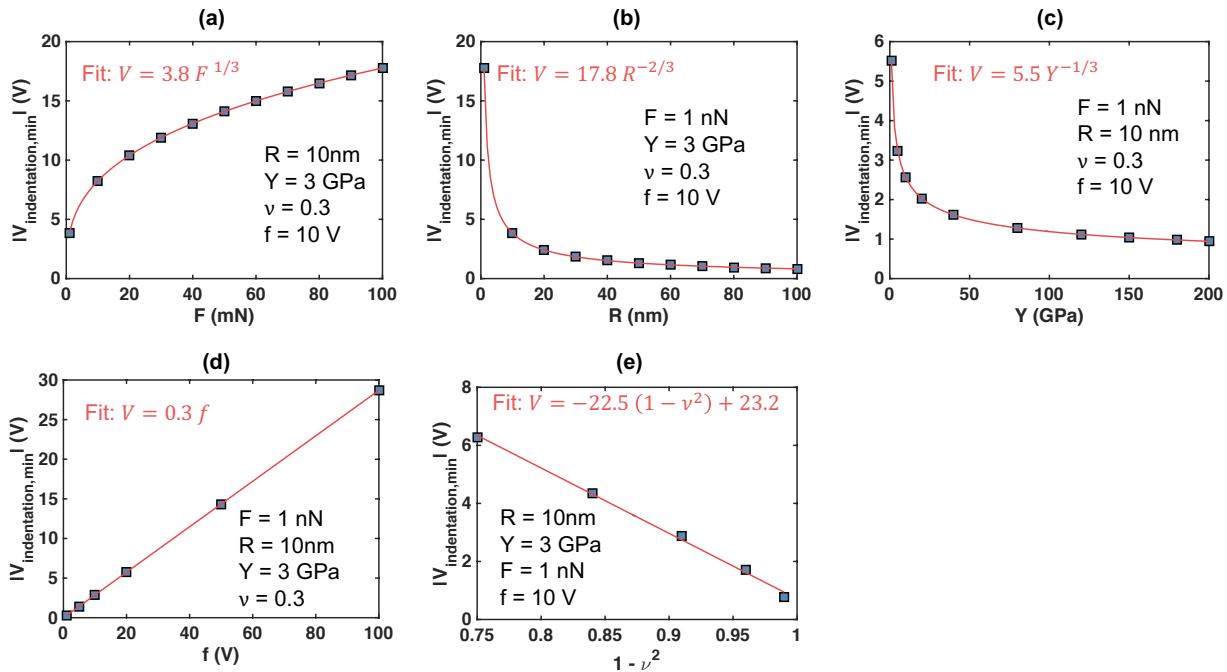

Fig. S2. Scaling of the magnitude of the minimum surface potential difference during indentation with applied force (F), indenter radius (R), Young's modulus (Y), flexocoupling voltage (f), and Poisson ratio ($\nu$). Surface potential differences are calculated numerically (blue squares) by varying one quantity while keeping all other parameters constant (constant values are black text in each plot). Red lines show fits to the calculated values and the equation of fit is in red text.

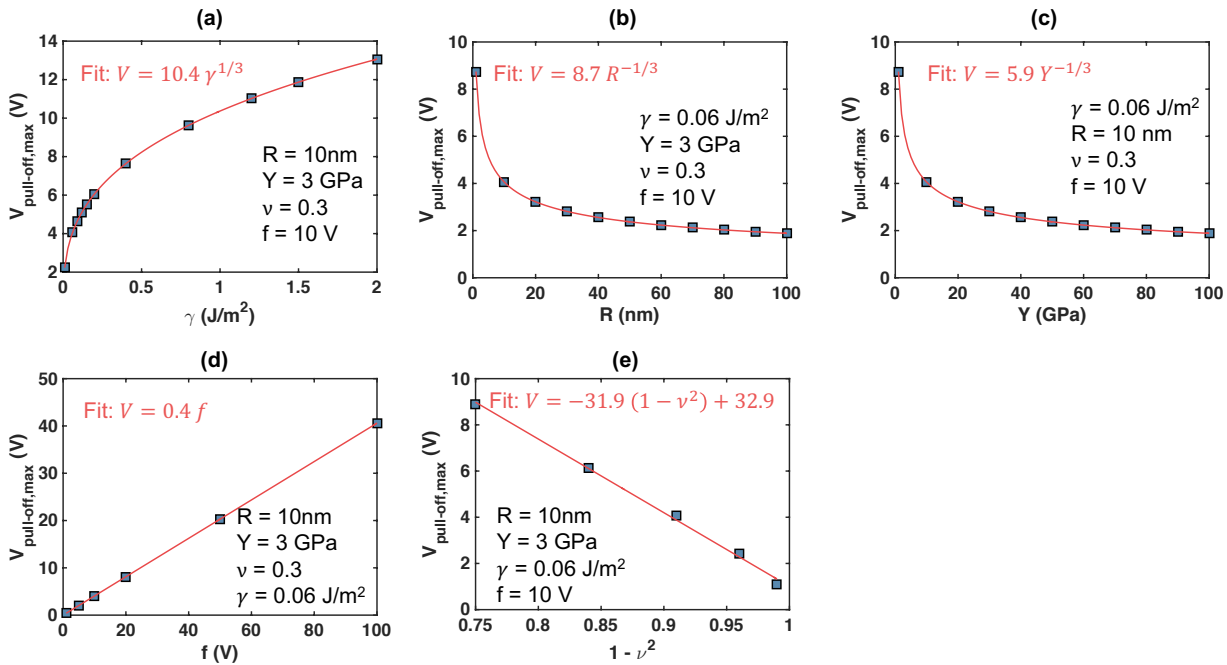

Fig. S3. Scaling of the magnitude of the maximum surface potential difference during pull-off with adhesion energy ($\Delta\gamma$), indenter radius (R), Young's modulus (Y), flexocoupling voltage (f), and Poisson ratio ($\nu$). Surface potential differences are calculated numerically (blue squares) by varying one quantity while keeping all other parameters constant (constant values are black text in each plot). Red lines show fits to the calculated values and the equation of fit is in red text.

## S8. Surface charge density and flexoelectric polarization

This model was developed under the assumption there is no free charge present to screen the polarization/electric field arising from strain gradients via a flexoelectric coupling. In reality, free charge will be present (e.g. from bulk defects, surface defects, or nearby air and water) and tend to accumulate on the surface of the deformed body to screen the polarization developed via the flexoelectric effect. Therefore, an estimate for upper bound of the surface charge density is the value of the flexoelectric polarization. In both the indentation and pull-off cases, the average polarization in the deformation volume is related to the average effective strain gradient in the deformation volume via

$$\overline{P}_z = \mu \left. \overline{\frac{\partial \epsilon}{\partial z}} \right|_{eff} \quad (S31)$$

As established in S5, $\left. \overline{\frac{\partial \epsilon}{\partial z}} \right|_{eff}$ is only a function of indenter size making $\overline{P}_z$ a function of indenter size and the flexoelectric coefficient $\mu$. Unfortunately, the flexoelectric coefficient is not a well-characterized materials property, so for this analysis a typical value of $\mu = 10^{-9}$ C/m is assumed [16].

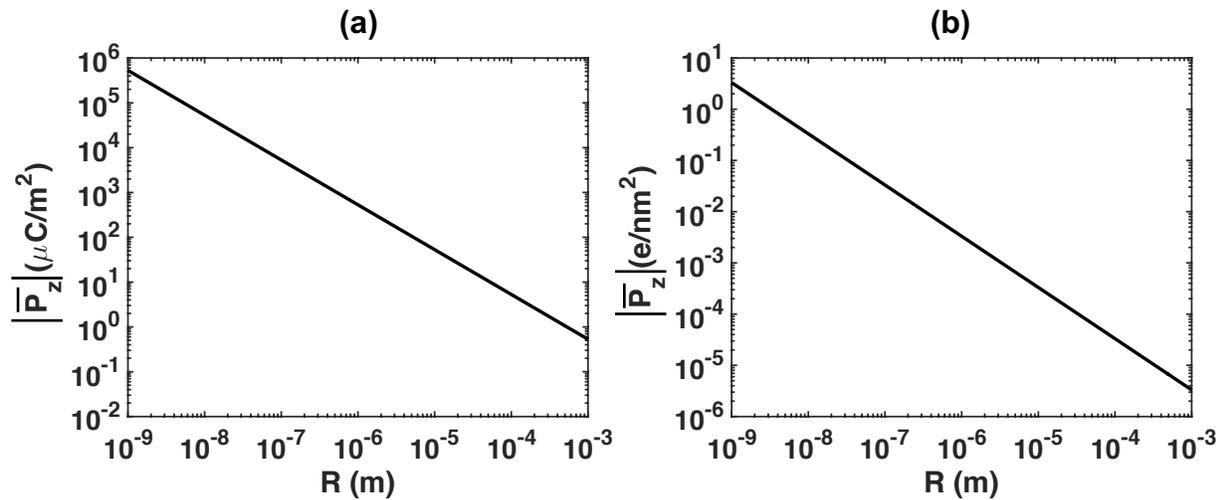

Fig. S4. Average flexoelectric polarization ($|\overline{P}_z|$) as a function of indenter size (R) assuming a flexoelectric coefficient of 1 nC/m. in units of $\mu C/m^2$ and $e/nm^2$.

**S9. Surface and interface contributions to the effective flexocoupling voltage**

A centrosymmetric material has no bulk piezoelectric terms, but the presence of a free surface or an interface breaks inversion symmetry which can lead to a piezoelectric contribution in the selvedge region near the surface. From elasticity theory this will decay exponentially into the material, perhaps extending 1-2 nm. This is relatively small compared to the typical size of the displacement field around an asperity, unless it is very sharp. Including these contributions will have a minor effect, but will not change the general results.

In addition to this, the physically measurable effective flexocoupling voltage is the sum of a bulk flexocoupling voltage (i.e. the intrinsic electronic and lattice response to a strain gradient deformation) and a surface/interface flexocoupling voltage (i.e. the change in the potential offset arising from a strain deformation) [40]. Unlike many properties, such as the piezoelectric effect, this surface term does not tend to zero in the limit of thick slabs [10,41,42]. In this work, an effective flexocoupling voltage is used to characterize the flexoelectric response of the deformed body. The microscopic details of the flexocoupling voltage do not change the results of our analysis because the magnitudes of both the bulk and surface contributions to the flexocoupling voltage are of the order of 1-10 V [41].

**S10. Experimental and theoretical ways to assess this model**

We will provide here some additional possibilities to both test and extend the model described herein, as well as advance further the science.

A very obvious piece of information that will be required as a prerequisite is flexocoupling voltage measurements for the materials used for triboelectric measurements. While there is now a small database of values for different materials, to firmly connect triboelectric and flexoelectric contributions there is a clear need for more measurements of flexoelectric coefficients in more materials, particularly as there may be subtle, unexpected contributions from, for instance, fillers in polymers. There is also a need for measurements in more complex and technologically relevant materials, for instance what is the flexocoupling voltage of cat hair or synthetic fibers in clothes?

Turning to specifics of our model, one of the most compelling pieces of support is the bipolar current measured during sliding experiments [43]. There are many possibilities to go beyond this to test details of our model and the underlying physics using scanning probe methods. As some examples:

1. Perform experiments where the tribocurrent/voltage only arises from normal force components with no shear.
2. Perform experiments where the tribocurrent/voltage is measured during pull-off. The elasticity problem is fairly well understood so the flexoelectric contributions can be calculated fairly well, and compared to experimental results.
3. Perform pull-off experiments as in 2. above, and combine this will Kelvin probe force microscopy to measure the surface potential changes.

Another interesting set of experiments, which will also connect to modelling would be to go beyond simple conical asperities to other cases where the elasticity problem of tribological contacts is well established. For instance, it would be informative to have triboelectric measurements between sinusoidal modulated surfaces or grids, or by using interlocking gears in micromechanical (MEMS) devices. In both cases it is in principle possible to simultaneously measure frictional forces, displacements and triboelectric currents/voltages, and cross-connect the two.

On the computational side, it should be possible to validate our model by extending existing flexoelectric phase-field models (e.g. [44-46]) or using finite element methods to conditions relevant to nanoscale asperity contact (e.g. [47]). This will be complicated by the need to explicitly account for charge-transfer [48]. Another set of in-silico expansions would be to consider the specific case of shear in more detail, as well as some of the other experimental samples mentioned above.

In an ideal world one would want to use a full ab-initio approaching using density functional theory. Recent developments of the first principles theory of flexoelectricity [8,9,12,49] have allowed for ab-initio calculations of bulk and surface [10,41] contributions to the flexoelectric response. At the time of writing the agreement between experimental results and these theoretical calculations is encouraging, but not yet good enough for reliability, particularly when it comes to the sign of the flexocoupling voltage. Hopefully in the near future this will improve. One could then couple ab-initio calculations with multiscale modelling to determine changes in interfacial charge density induced by strain/strain gradient deformations and thereby quantify the importance of flexoelectricity in triboelectric (and other tribological) phenomena.